\documentclass[useAMS]{mn2e}
\usepackage{epsfig}
\usepackage{graphicx}
\usepackage{amsmath, amssymb}

\title[Planetary Migration to Large Radii]
{Planetary Migration to Large Radii}
\author[R.G. Martin, S.H.Lubow, J.E. Pringle \& M C. Wyatt]
{R.G. Martin$^{1}$, S.H. Lubow$^2$,  J.E. Pringle$^{1,2}$ \& M.C. Wyatt$^1$
\\
$^1$Institute of Astronomy, Madingley Road, Cambridge, CB3 0HA, UK\\
$^2$STScI, 3700 San Martin Drive, Baltimore, MD 21218, USA\\ }

\begin{document}

\maketitle
\begin{abstract}
  There is evidence for the existence of massive planets at orbital
  radii of several hundred AU from their parent stars where the
  timescale for planet formation by core accretion is longer than the
  disc lifetime. These planets could have formed close to their star
  and then migrated outwards. We consider how the transfer of angular
  momentum by viscous disc interactions from a massive inner planet
  could cause significant outward migration of a smaller outer planet.
  We find that it is {\it in principle} possible for planets to
  migrate to large radii. We note, however, a number of effects which
  may render the process somewhat problematic.
\end{abstract}

\begin{keywords} accretion,accretion discs - planets and
  satellites: formation - planetary systems: protoplanetary
  discs.
\end{keywords}

\section{Introduction}

Most stars at an age of about $10^6\,\rm yr$ are surrounded by
circumstellar discs which are optically thick at optical and infrared
wavelengths (Strom et al. 1989, Kenyon \& Hartmann 1995, Haisch, Lada
\& Lada 2001). At earlier times, since these discs are part of the
star formation process, it seems likely that they contained a
significant fraction of the stellar mass (e.g. Lin \& Pringle, 1976).
At this age of around $10^6\,\rm yr$, the discs are found to be as
massive as a few percent of a solar mass and have typical sizes of a
few hundred AU (Beckwith et al.  1990). It must be within these discs
that planets form.  Most of the detections of extrasolar planets are
based on Doppler techniques and a smaller number of detections come
from occultations \footnote{See http://exoplanet.eu for a recent
  catalogue of exoplanets.} (Butler et al. 2006). For obvious reasons
these detection methods are biased toward finding planets at
relatively small distances from their parent stars compared to the
typical disc sizes, and compared to the planets on the Solar System
(e.g. Beer et al., 2004). In addition the planets discovered are
almost all high mass, and therefore most likely, all gas-rich giants.
Current ideas on the formation of such planets, involving the initial
formation of a metal-rich core and subsequent accretion of a gaseous
envelope from the disc, suggest that these planets formed at larger
radii (around a few AU so that core-formation can occur) and then
migrated inwards to their current positions (e.g., Lin, Bodenheimer,
\& Richardson 1996).  However, in addition to these planets at small
radii, there is evidence, albeit somewhat indirect, for several
massive planets at very large radii, much greater than the putative
formation radius of around a few AU.  For example, one model for the
formation of spiral structure seen in the disc of HD141569 involves
the presence of a planet of mass $0.2-2\,\rm M_J$ orbiting at
$235-250\,\rm AU$ and a Saturn-mass planet at $150\,\rm AU$ (Wyatt
2005a). Gaseous discs appear to dissipate on a timescale of $5-10\,\rm
Myr$ (Haisch, Lada \& Lada 2001) and clearly the gas giant planets
must form before this occurs. In the standard core accretion models
(Safronov 1969), the timescale for building a giant planet core scales
with the square of the distance from the central star (Pollack et al.
1996). The formation time for Jupiter (at $5\,\rm AU$) is estimated to
be $1-10\,\rm Myr$ .  Hence, to form a $1\,\rm M_J$ planet at
$200\,\rm AU$ would take $1.6-16\,$Gyr, by which time the gaseous disc
has long since disappeared.  It is therefore thought highly unlikely
that gas giant planets can form {\it in situ} at large distances from
the central star.

One obvious possible explanation for the presence of such planets at
large radii is that they form at a standard radius and migrate
outwards.  The key issue is then what provides the outward torque.
Inward migration seems a more natural possibility, since the disc,
which facilitates the migration, is part of the accretion process.
For example, consider a planet whose mass is small compared to the
disc but sufficiently large to open a gap in the disc. Such a planet
follows the radial mass flow of the disc in what is commonly called
Type II migration (Lin \& Papaloizou 1986).  Generally, disc material
flows inward as an accretion disc carrying the planet inward towards
the central star. However, in order for accretion to proceed, the disc
must move angular momentum outwards. Thus some material in the outer
parts of an accretion disc moves outwards in order to conserve the
angular momentum lost by most of the disc. The fraction of outward
moving material decreases with time. A planet located in such a disc
region would similarly move outward, at least for a limited
time. Veras \& Armitage (2004) considered this possibility. They model
a situation in which disc photoevaporation at large radii causes a
planet at (small) standard radii to effectively experience outer edge
effects which in turn causes outward migration.  For the models they
considered, planets were able to migrate to about $20\,\rm AU$ and in
a few cases as far as $50\,\rm AU$.

There are two main problems to be addressed in trying to drive a
planet to large distance. First it is necessary to provide a large
enough source of angular momentum. And second, it is necessary to give
the planet that angular momentum on a short enough timescale. In the
models of Veras \& Armitage (2004) the angular momentum was provided
by the disc. Here we build on these ideas, but consider the
possibility that the source of the angular momentum is an inner, more
massive, planet.  As a simple example, suppose initially we have two
planets of masses $M_1$ and $M_2$ orbiting a star in circular orbits
at radii $a_{1}$ and $a_{2}$ respectively with $a_{2}>a_{1}$. Then if
the inner planet could somehow be induced to give all its angular
momentum to the outer one, the outer one would move out to a final
radius
\begin{equation}
a_f = \left( \frac{M_1}{M_2} a_1^{1/2} + a_2^{1/2} \right)^2
\label{simple}
\end{equation} 
where we have assumed, as is reasonable if the angular momentum
transfer is effected through tidal interactions with a gaseous disc,
that the outer planet remains in a circular orbit.  Suppose we have a
planet of mass $5\,\rm M_J$ that begins at $a_1 = 5\, \rm AU$ and a
planet of mass $1\,\rm M_J$ that begins at $a_2 = 10\,\rm AU$.  If the
inner planet manages to give up all its angular momentum, and so ends
up at the central star with radius $R \ll a_1$, then the outer planet
can migrate to a distance $a_{\rm f} = 206\,\rm AU$.  This
demonstrates that mutual interaction within a two (or more) planet
system can, {\it in principle}, drive a planet out to the required
distances of a few hundred AU. 

In this paper we investigate what is required in practice for this to
be achieved.  It is clear that there are two main requirements. First
we need some efficient mechanism for transferring angular momentum
from the inner planet to the outer one. Second, we need to effect the
transfer on a sufficiently short timescale before the means of
transferring the angular momentum, presumably tied to the disc, has
vanished.

The outline of the paper is as follows.  In Section~2 we consider the
evolution of a region of a steady accretion disc that lies exterior to
a newly formed massive planet or companion star. We show that the disc
changes format from accretion to decretion and in doing so changes
its surface density profile on a viscous timescale. In Section~3 we
consider the evolution of a steady accretion disc in which two planets
are permitted to form and consider the circumstances necessary for
the outer planet to be forced to migrate outwards to large radius.  By
an age of $10^7\,\rm yr$ it appears that massive circumstellar discs
are no longer present. All that remains is possibly a debris disc.
Photoevaporation of the disc is considered a possible gas dispersion
mechanism. In Section~4 we investigate the effects of mass loss from
the disc caused by photoevaporation. In Section~5 we discuss the
applicability of our findings to some observed systems. We summarise
our conclusions in Section~6.

\section{The change from accretion disc to decretion disc}
\label{sec:addec}

An accretion disc occurs when the disc has a mass sink at its inner
edge (allowing accretion onto the central object) while providing no
torque at the inner edge. A decretion disc occurs when the central
object does not accrete, but instead provides a central torque which
prevents inflow. For a decretion disc, the torque exerted by the
central object results in radial outflow.  Decretion discs can arise
when the central torque is provided by a binary star or star-planet
system. We consider in this Section how an accretion disc adjusts when
conditions at the inner boundary change so that accretion there no
longer occurs. We consider the idealised situation in which, for
example, a massive planet, or binary companion, forms suddenly
(compared to a viscous timescale) near the inner disc edge. Then the
tidal influence of the inner binary is assumed to prevent further
accretion, and to provide the disc with angular momentum through a
tidal torque. What we are interested in here is the resulting change
to the disc structure caused by such an event.

The equation that describes the evolution of a flat Keplerian
accretion disc with surface density $\Sigma(R,t)$ where $R$ is the
radius from centre of the star and $t$ is the time, is
\begin{equation}
\frac{\partial \Sigma}{\partial t}=\frac{1}{R}\frac{\partial}{\partial R}\left[3R^{1/2} 
\frac{\partial}{\partial R}(\nu \Sigma R^{1/2})\right]
\label{first}
\end{equation}
(Pringle 1981) where $\nu (R,\Sigma)$ is the kinematic viscosity.  In
general we shall model the effect of the planet or companion star on
the disc as an extra angular momentum source term. The governing
equation for such a disc is
\begin{equation}
\frac{\partial \Sigma}{\partial t}=\frac{1}{R}\frac{\partial}{\partial R}\left[3R^{1/2} 
\frac{\partial}{\partial R}(\nu \Sigma R^{1/2})
-\frac{2\Lambda \Sigma R^{3/2}}{(GM_{\star})^{1/2}}\right],
\label{lambda}
\end{equation}
where $\Lambda (R,t)$ is the rate of input of angular momentum per
unit mass (Lin \& Papaloizou 1986, Armitage et al. 2002).  We analyse
here the evolution of disc material that resides outside the orbit of
the companion.

In this Section, for the purposes of illustration, the companion is
assumed sufficiently massive that its angular momentum is much greater
than that of the disc, so that the orbital evolution of the central
objects can be ignored. We consider a high mass planet or companion
star which orbits at fixed radius and prevents mass from passing
interior to the disc inner edge.  In this situation, the effect of
this extra torque term involving $\Lambda$ is equivalent to imposing a
zero radial velocity inner boundary condition at $R=R_{\rm in}$
(Pringle 1991) where the radial velocity in the disc is given by
\begin{equation}
V_{\rm R}=-\frac{3}{\Sigma R^{1/2}}\frac{\partial}{\partial R}(R^{1/2}\nu \Sigma).
\end{equation}

We initially take an accretion disc with constant accretion rate but
truncated at some outer radius $R_{\rm t}$. Thus for the initial
surface density we take
\begin{equation}
\label{sd}
\Sigma (R)=\Sigma_0 \left(\frac{R_{\rm in} }{R}\right)^{\beta}
\end{equation}
in the range $R_{\rm in}<R<R_{\rm t}$ where $\Sigma_0$ is a constant
which determines the initial mass of the disc and $R_{\rm t}$ is the
initial outer edge of the disc.  We take $\beta=3/2$ (Weidenschilling
1977, Hayashi 1981) in most of our numerical simulations, but also
investigate the effect of varying $\beta$.

In order that $\nu \Sigma = const.$ in the initial accretion disc, and
to keep things simple at later times, we choose the kinematic
viscosity of the disc to be
\begin{equation}
\label{viscosity}
\nu = \nu_0 \left(\frac{R}{1\,{\rm AU}}\right)^{\beta}
\end{equation}
with $\nu_0=2.466\times 10^{-6}\, \rm AU^2\,yr^{-1}$ to give a
reasonable time scale for the evolution of the disc of a few Myr.  In
terms of the $\alpha$-prescription for the viscosity $\nu=\alpha
c_{\rm s}H$ where $\alpha$ is dimensionless, $c_{\rm s}$ is the sound
speed and $H$ is the scale height of the disc, and for a typical value
$H/R = 0.05$ (see below), this implies
\begin{equation}
\label{alphavalue}
\alpha=1.57\times10^{-4}\left(\frac{R}{\rm AU}\right)^{(\beta-1/2)} .  
\end{equation}

We solve equation~(\ref{lambda}) on a fixed, uniform mesh in the
variable $x=R^{1/2}$ by using a simple first order explicit numerical
method.  We use $4000$ grid points with a zero radial velocity inner
boundary condition at $R_{\rm in}=5\,\rm AU$, $V_{\rm R}(R_{\rm in},t)
=0$ for all time $t$ and a zero torque outer boundary at $R_{\rm
  out}=1500\,\rm AU$.  Due to the inner boundary condition on the
radial velocity, the gravitational torque term $\Lambda$ in
equation~(\ref{lambda}) is ignored.  The outer boundary condition does
not effect the disc evolution because it is well outside the disc over
the course of the evolution we consider.

\begin{figure}
\epsfxsize=8.4cm 
\epsfbox{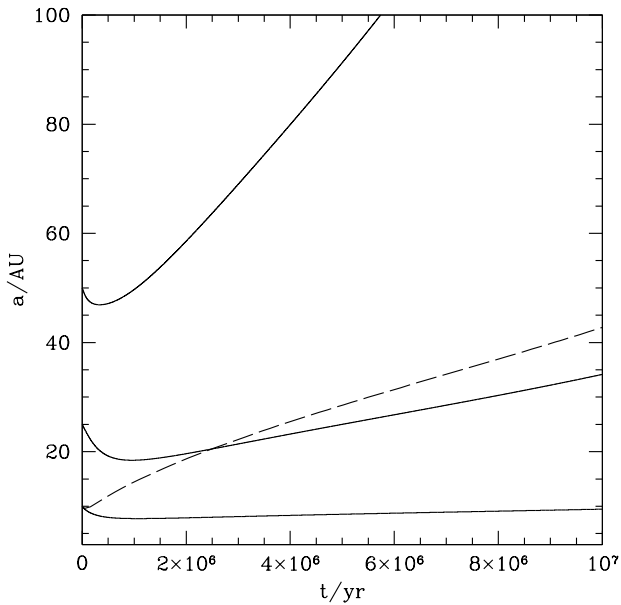}
\caption[] { Particle paths as a disc changes from accretion to decretion.
  The vertical axis is the particle orbital radius in AU and the
  horizontal axis is the time in years. The disc viscosity is given by
  equation~(\ref{viscosity}) with $\beta = 3/2$ and the disc inner
  edge is located at $R_{in}=5\,\rm AU$.  At time $t=0$ the disc is taken
  to be a steady accretion disc. The solid lines are for paths located
  in a disc that is initially truncated at $R_{\rm t}=100\,\rm AU$. The
  dashed line is for a path in a disc with $R_{\rm t}=20\,\rm AU$.  The
  effect of a central torque is provide by the boundary condition on
  the radial velocity $V_{\rm R}=0$ at the disc inner edge. }
\label{trace}
\end{figure}

In the simulations, we trace the radial movement of disc particles by
integrating the radial velocity because
\begin{equation}
\frac{da}{dt} = V_{\rm R}(a(t), t),
\label{dadt0}
\end{equation}
where $a(t)$ is the radial position of the particle. The imposition of
a non-absorbing boundary at an inner radius $R_{\rm in}$ transforms
the initial accretion disc into a decretion disc. To show the outcome
of imposing such a boundary condition, we plot in Figure~\ref{trace}
the radial motions of particles in two discs, one initially truncated
at $R_{\rm t}=20\,\rm AU$ (dashed line in Figure~\ref{trace}) and the
other initially truncated at $R_{\rm t}=100\,\rm AU$ (solid lines in
Figure~\ref{trace}). In both discs, the particles initially follow the
inward accretion, before reversing and being expelled to large radii.
The smaller disc makes the initial adjustment more quickly and expels
the particles more quickly than the larger disc.

To understand the disc flow analytically, we consider the behaviour of
self-similar solutions to the disc flow that are valid at large times
based on Pringle (1991).  We first consider the case of an accretion
disc.  In a steady accretion disc, the left-hand side of equation
(\ref{first}) is zero. This condition is satisfied for
\begin{equation}
V_{\rm R} = -\frac{3 \nu}{2R}
\label{VRaSS}
\end{equation}
and the 
mass flux is given by
\begin{equation}
\dot{M} = -3 \pi \nu \Sigma
\label{MFSS}
\end{equation}
which is independent of radius. The latter provides the radial
variation $\Sigma(R)$ in the steady-state.  Over time, an arbitrary
initial disc which evolves with absorbing (mass flow) inner boundary
conditions will approach this steady-state accretion disc.  The
self-similar solution valid at large times for an accretion disc
implies that
\begin{equation}
V_{\rm R}= -\frac{\nu_0}{R_0} \left (\frac{3 r^{\beta-1}}{2} - \frac{r }{(2-\beta) \tau} \right),
\label{VRalt}
\end{equation}
where $\nu_0$ is given by equation~(\ref{viscosity}) with $R_0=1\,\rm
AU$, dimensionless radius $r=R/R_0$ and dimensionless time $\tau = t
\nu_0/R_0^2$.  Consequently, at fixed radius and for sufficiently
large $t$, the radial velocity in equation~(\ref{VRalt}) approaches
the steady state value given in equation~(\ref{VRaSS}) which provides
the Type II migration velocity and the motion is always eventually
inward.

In contrast, a steady decretion disc has $V_{\rm R}=0$ everywhere and
\begin{equation}
\nu \Sigma \propto R^{-1/2}
\label{SigS}
\end{equation} (Pringle 1991). 
The decretion disc simulations in Figure~\ref{trace} start with a
steady-state accretion disc density profile.  Initially, the disc is
unaware of the inner boundary where the outward torque is exerted,
and behaves as a standard accretion disc. Over time the mass builds up
near the inner boundary as a consequence of the blocked inflow. The
outward torque increases and its effects propagate outward.  Over
time, more of the disc material flows radially outward to become a
decretion disc. The disc gains angular momentum at the expense of the
central system.

We discuss below whether/how a decretion disc approaches its
steady-state density and velocity over time. We also determine the
particle drift velocity $V_{\rm R}$ for a decretion disc and thereby
obtain its Type II migration rate.  It can be shown from the
self-similar solutions for a decretion disc (using the equations of
Pringle (1991) for a viscosity which is non-linear in $\Sigma$ and
applying the limit of the $\Sigma$ variation disappearing) that the
density and velocity evolves at large times as
\begin{equation}
\Sigma(r,\tau) =\frac{ \Sigma_0 \exp[- r^{2-\beta}/(b\, \tau)] }{\tau^c r^{\beta+1/2}}
\label{densa}
\end{equation}
and
\begin{equation}
V_{\rm R}(r,\tau) =  \frac{\nu_0}{R_0}\frac{r}{(2-\beta) \tau},
\label{VRae}
\end{equation}
where $\nu_0$, $R_0$, $r$ and $\tau$ are as defined below equation
(\ref{VRalt}), $\Sigma_0$ is the density normalisation constant,
$b=3(2-\beta)^2$ and $c=1-1/(4-2\beta)$. This result shows that for
fixed $r$ the density at large time varies approximately as a power
law in time.  At fixed $r$ and large $\tau$ the density decreases in
time for $\beta<3/2$, remains constant in time for $\beta=3/2$ and
increases in time for $3/2<\beta<2$ as a consequence of the mass
flux variation in radius.  Therefore, the density approaches a nonzero
steady-state value only for $\beta=3/2$.  For $0<\beta<2$, the spatial
variation of the density at large fixed time does satisfy
equation~(\ref{SigS}) where the proportionality constant in that
equation generally depends on time.  For $0<\beta<2$, at fixed radius
the radial velocity $V_{\rm R}$ at all radii approaches the
steady-state value of zero at large time.
 
The particle trajectory $a(t)$ at large times is determined by
applying equation~(\ref{VRae}) in equation~(\ref{dadt0}).  We find
that
\begin{equation}
a(\tau)=d \, \tau^{1/(2-\beta)}
\label{Ra}
\end{equation}
and
\begin{equation}
\frac{d a}{d t}= V_{\rm R}(a(t), t) =  \frac{\nu_0\, d}{R_0^2}
\frac{\tau^{\frac{\beta-1}{2-\beta}}}{2-\beta},
\label{VRa}
\end{equation}
where $d$ is a constant with units of length along each particle path
which is determined by its initial radius and radial velocity.
Equation~(\ref{Ra}) shows that the evolution at large times proceeds
as the particles in general move to large radii for $0<\beta<2$.
However if $d=0$, the path remains fixed at small radii.  This
situation occurs for example at the disc inner boundary. Even small
values of $d$ eventually lead to outflow to large distances if the
disc survives for a sufficiently long time.

Equation~(\ref{VRa}) provides the Type II migration velocity in a
decretion disc.  Notice that $V_{\rm R}$ goes to zero at large time for
$0<\beta<1$, is constant for $\beta=1$ and increases in time for
$1<\beta<2$.  There is a marked difference in behaviour between the
(Lagrangian) particle velocity and the (Eulerian) velocity at fixed
$R$, as discussed earlier which always vanishes at large $t$. The
reason for the difference is that a particle having $d>0$ will never
experience viscously relaxed conditions as it moves outwards while
the velocity at fixed radius declines to zero over several local
viscous times.  This is a consequence of the particle velocity being
of order the characteristic viscous propagation speed $ \sim
\nu(a(t))/a(t)$ for $d \sim R_0$.

The computations described in this Section can be taken as the
limiting case of what happens if a stellar companion or massive planet
is introduced and truncates the outer disc at an inner radius of
$5\,\rm AU$.  The paths traced by the particles represent the radial
motion of a light planet undergoing Type II migration in the outer
disc. In this case direction of motion of the light planet is reversed
on the local viscous timescale and the planet can eventually be
expelled to a large radius by the presence of the inner objects.  The
simplifying assumptions made in this Section imply that there is an
infinite source of angular momentum at the inner disc edge. In the
next Section we describe a more realistic calculation with the two
planets and the disc all of finite and comparable, mass.

\section{Two Planets}

In this section we consider the migration of two planets in the disc.
We shall take the inner planet to have mass $M_1 = 5 \,\rm M_J$ and
the outer planet to have mass $M_2 = 1 \,\rm M_J$. We use a model of
planetary migration in a disc similar to that of Armitage et al.
(2002). This is a one-dimensional model evolving by internal viscous
torques and external torques from planets embedded in the disc
(Goldreich \& Tremaine 1980, Lin \& Papaloizou 1986, Trilling et al.
1998, Trilling, Lunine \& Benz 2002).   We consider only planets
massive enough to open a clean gap, i.e. pure type II migration, and
cases where the planets are sufficiently well separated that there is
gas between them. If they are too close, this is not the case and the
evolution proceeds in a different manner from that found in this paper
(e.g. Kley, Peitz \& Bryden, 2004). 

The criterion given by Lin \& Papaloizou (1986) for the opening of a
clean gap, and so for the validity of simple Type II migration, can be
written in terms of the local Reynold's number $Re = R^2 \Omega /\nu$,
where $R$ is the radius of the planet's orbit and $\Omega$ the angular
velocity of the disc flow there. The condition for gap opening is that
\begin{equation}
\label{LP86}
Re \ge 40 \left( \frac{M}{M_p} \right)^2 \left( \frac{H}{R} \right)^3,
\end{equation}
where $M$ is the mass of the star and $M_{\,\rm p}$ the mass of the
planet.  The detailed protoplanetary disc models of Bell et al. (1997)
suggest that to a reasonable approximation we may assume $H/R = 0.05$,
and so we make this assumption throughout. For a planet mass $M_{\rm
  p} =1 \,\rm M_J$ and for this value of $H/R$ this implies that Type
II migration occurs when $Re \ge 5.5 \times 10^3$.  If $\nu = \alpha c_s
H$ then the Reynold's number can be written as
\begin{equation}
Re = \left( \frac{R}{H} \right)^2 \frac{1}{\alpha}. 
\end{equation}
For the value of $\alpha$ given by equation~\ref{alphavalue} with
$\beta=3/2$, we find $Re = 2.55 \times 10^6 (R/ \rm{AU})^{-1}$. If
these estimates hold, then we conclude that we may safely assume
simple Type II migration for the calculations in this paper.

In practice, there may be some flow through the gap when full 2D or 3D
effects are taken into account in the simulations (Artymowicz \&
Lubow, 1996; Lubow \& D'Angelo, 2006). D'Angelo, Lubow \& Bate (2006)
find that for planets on circular orbits, and with $Re \approx 10^5$,
the 1D migration rate agrees to within a few percent with 2D
simulations for planets whose masses are of order $1\,\rm M_J$ and
greater.

Recently Crida, Morbidelli \& Masset (2007) and Crida \& Morbidelli
(2007) have pioneered a more sophisticated approach which involves
modelling the gas flow close to the planet using an evolving 2D grid,
combined with the 1D approach for the rest of the disc. For a planet
with mass $M_p = 1\,\rm M_J$ and a disc with $H/R \approx 0.05$, they
find that clean Type II migration occurs only when the local
Reynold's number exceeds $10^5$. For Reynold's numbers $10^4 \le Re
\le 10^5$ the torque felt by the planet is  reduced  relative to the
usual Type II torque by a factor of order $Re/10^5$. For lower
Reynold's numbers the torque is found to reverse.

In this paper we are interested in the migration of a nominal $1\,\rm
M_J$ planet from around $10\,\rm AU$ to $200\,\rm AU$, for which our
Reynold's numbers are in the range $2.5 \times 10^5$ to $10^4$. If the
work of Crida \& Morbidelli (2007) is correct, then the results given
here overestimate the efficiency of outward migration by as much as an
order of magnitude. However, since the critical Reynold's number
varies as the square of planetary mass (equation~\ref{LP86}), the
results, suitably scaled, would still be valid for the outward
migration of a planet of $\sim 3\,\rm M_J$.

In view of the uncertainty of the actual time-dependent properties of
planet forming discs, in particular with regard to disc thickness and
magnitude of viscosity in comparison to the simple formulae employed
here, in the following we shall simply use the standard Type II
migration torque formulae. We shall find that the conditions under
which outward migration can be achieved are limited, and it therefore
needs to be borne in mind that they may in fact be more limited still.

In this case, the governing equation becomes

\begin{align}
\frac{\partial \Sigma}{\partial t}=&\frac{1}{R}\frac{\partial}{\partial R}\left[3R^{1/2} 
\frac{\partial}{\partial R}(\nu \Sigma R^{1/2}) \right]\cr
&-\frac{1}{R}\frac{\partial}{\partial R}\left[\frac{2(\sum_{i=1}^2 \Lambda_i) \Sigma R^{3/2}}{(GM_{\star})^{1/2}}\right]\
-\dot \Sigma_{\rm w}.
\label{maineq}
\end{align}
where $\Lambda_i(R,a_i)$, for $i=1,2$, is now the rate of angular momentum
transfer per unit mass from planet $i$ to the disc where $a_i(t)$ is
the distance of the planet from the star of mass $M_{\star}$.  In
Section \ref{sec:photo} we shall consider the possibility of mass loss
from the disc caused by a wind, represented by the term $\dot
\Sigma_{\rm w}(R,t)$ on the right-hand side of the equation.  In this
Section we consider the case for $\dot \Sigma_{\rm w}(R,t)=0$.

We take the torque distribution to be of the form
\begin{align}
\Lambda_i (R,a_i)=\begin{cases}\displaystyle{-\frac{q_i^2GM_\star}{2R}\left(
\frac{R}{\Delta _{\rm p}}\right)^4}& \text{if $R \le a_i$}, \cr
\displaystyle{\phantom{-}\frac{q_i^2GM_\star}{2R}\left(
\frac{a_i}{\Delta _{\rm p}}\right)^4}&  \text{if $R \ge a_i$},\end{cases}
\end{align}
(Armitage et al. 2002) where $q_i=M_{i}/M_{\star}$ is the ratio
between the mass $M_{i}$ of planet $i$ and that of the star and
\begin{equation}
\Delta _{\rm p} = \max (H,|R-a|),
\end{equation}
where $H$ is the scale height of the disc. As remarked above, we
assume $H=0.05R$.  By Newton's third Law, the orbital migration of
planet $i$ occurs at a rate
\begin{equation}
\frac{da_i}{dt}=-\left(\frac{a_i}{GM_{\star}}\right)^{1/2}\left(\frac{4\pi}{M_{{\rm p}i}}\right)
\int_{R_{\rm in}}^{R_{\rm out}}\!\!\!\!\!\!\!  \Lambda_i \Sigma\, R\,dR,
\label{dadt}
\end{equation}
(Lin \& Papaloizou 1986). Note that we neglect the gravitational
interactions between the two planets.

As before, we solve equation~(\ref{maineq}) on a fixed mesh which is
uniform in the variable $x=R^{1/2}$, by using a simple first order
explicit numerical method. We use 4000 grid points with an inner
boundary at $R_{\rm in}=0.01\,\rm AU$ and an outer boundary at $R_{\rm
  out}=900\,\rm AU$.  We take zero torque boundary conditions at both
$R_{\rm in}$ and $R_{\rm out}$.  The outer boundary is large enough to
not affect the disc.  The inner boundary condition allows all material
arriving there to be accreted by the central star.  Because the
viscous torque is given by
\begin{align} \notag
G&=-2\pi\nu\Sigma R^3 \frac{d\Omega}{dR} \\
&=\phantom{-}3\pi\nu\Sigma(GM_\star R)^{1/2},
\label{torque}
\end{align}
assuming $\Omega=(GM/R^3)^{1/2}$ for a Keplerian disc, the boundary
conditions are implemented by taking $\Sigma = 0$ at $R_{\rm in}$ and
at $R_{\rm out}$.

We assume that the planets are initially located at radii $a_1 =
5\,\rm AU$ and $a_2 = 10\,\rm AU$ and that the viscosity is described
by equation (\ref{viscosity}) with $\beta$ generally equal to $3/2$.
Because our interest is in discovering what conditions are required to
enable a sufficient degree of outward migration, we adopt a highly
simplified initial disc density distribution and gap sizes which we
describe below.

Over time, the disc gap sizes in the simulations adjust,
since they are determined dynamically by a competition between tidal
and viscous torques.  In each of the three ranges of radii, $(R_{\rm
  in}, a_1)$, $(a_1, a_2)$ and $(a_2, R_{\rm t})$, we assume that the
initial surface density profile is as given by equation~(\ref{sd}),
except that we may adjust the amount of mass in each region by setting
the local value of $\Sigma_0$.  We consider a form of the initial
density distribution that permits the specification of the disc mass
interior to the planets (inner disc), between the planets
(between-planet disc) and exterior to the outer planet (outer disc).

We take the initial surface density distribution to be of the form
\begin{equation}
\Sigma_0 = \begin{cases}
\displaystyle{\Sigma_{0 \rm in}} & \text{if $R_{\rm in}\le R \le a_1-\Delta a_{1-}$}, \cr
\displaystyle{0 }                  & \text{if $a_1-\Delta a_{1-} < R < a_1+\Delta a_{1+}$} \cr
\displaystyle{\Sigma_{0 \rm bet}} & \text{if $a_1+\Delta a_{1+} \le R \le a_2-\Delta a_{2-}$} \cr
\displaystyle{0  }                 & \text{if $a_2-\Delta a_{2-} < R < a_2+\Delta a_{2+}$}\cr
\displaystyle{\Sigma_{0 \rm out }} &  \text{if $a_2+\Delta a_{2+}\le R \le R_{\rm t}$}. \end{cases}
\label{densdistr}
\end{equation}
Here the gap widths $\Delta a_{i\pm}$ are chosen so that the mass
removed from the disc (inner, intermediate, or outer discs) to form
the gap is equal to the half mass of corresponding planet, provided
the disc mass is nonzero.  We shall find below only in some
circumstances can sufficient outward migration be achieved. For this
reason we do not assume the surface density is the same across gaps.
In addition, this disc is initially truncated at some radius $R_{\rm t}
\ll R_{\rm out}$.

\subsection{Disc only between the planets}

\begin{figure}
\epsfxsize=8.4cm 
\epsfbox{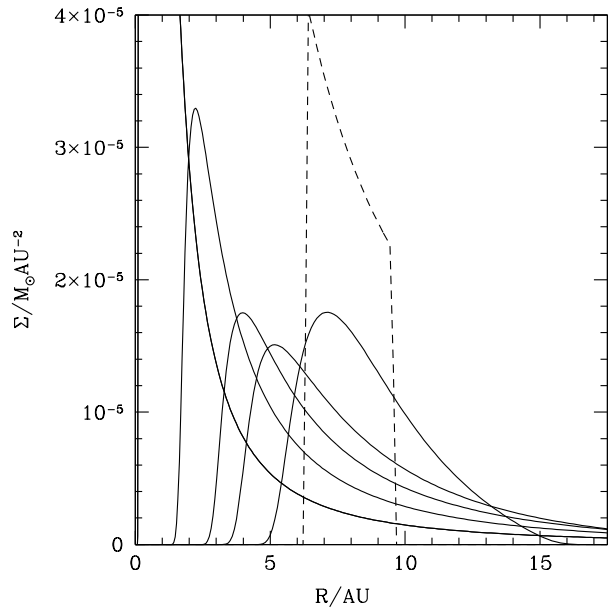}
\caption[] {Plot of disc surface density as a function of radius.
  At time $t=0$ there is a $5\,\rm M_J$ planet at $a_1 = 5\,\rm AU$
  and a $1 \,\rm M_J$ planet at $10\,\rm AU$. The gaseous disc has
  $\beta=3/2$ and consists of $5\,\rm M_J$ distributed only between
  the planets.  The density distribution at $t=0$ is shown by the
  dashed line. The subsequent evolution of the surface density is
  shown at times $4.0\times 10^4$, $1.4\times 10^5$, $2.5\times 10^5$,
  $5.5\times 10^5$ and $1.8\times 10^6\,\rm yr$ with the peak of the
  distribution moving monotonically inwards. }
\label{surf}
\end{figure}

Disc mass that is located outside a planet's orbit provides a negative
torque on the planet which pushes it inwards. The higher the outer
disc mass the stronger the torque and so the faster the inward
migration.  Similarly, disc mass located inside a planet's orbit
creates a positive torque which pushes it outwards. A minimum
requirement for the model under consideration here is that the inner
planet can transfer angular momentum to the outer one. For this to be
able to happen there must be disc mass between the planets.  We
consider first what happens if there is only a disc between the two
planets, i.e., $\Sigma_{0 \rm in}= \Sigma_{0 \rm out}=0$ in equation
(\ref{densdistr}).  In Figure~\ref{surf} we plot the surface density
evolution for the system with two planets, a $5 \,\rm M_J$ inner
planet and a $1 \,\rm M_J$ outer planet with a $5 \,\rm M_J$ gaseous
disc between them.  As seen in Fig~\ref{surf}, the peak of the density
distribution moves inwards in time and simultaneously the density
spreads outward. Since the disc lies at all times between the two
planets, it is evident that as the inner one moves inwards, the outer
one moves outwards.

\begin{figure}
\epsfxsize=8.4cm 
\epsfbox{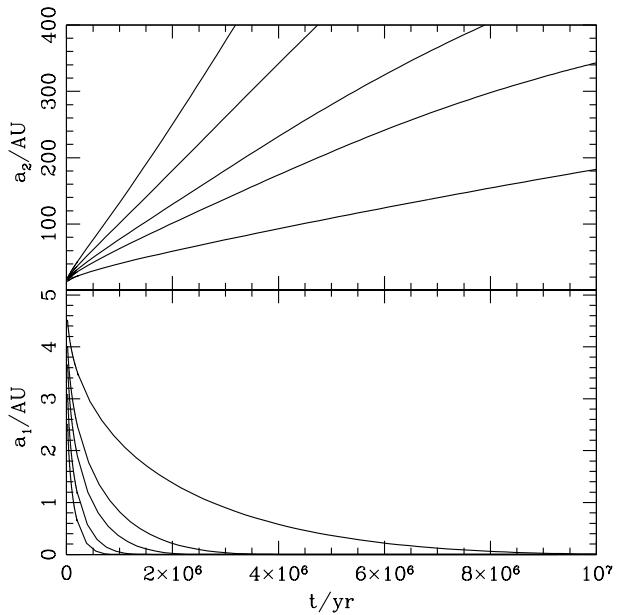}
\caption[] {Migration of the planets as a function of time for the
  case when the disc mass is distributed only between them according
  to equation~(\ref{densdistr}) with $\Sigma_{0 \rm in } = \Sigma_{0
    \rm out } = 0$ . The initial configuration of the planets is as
  described for Figure~\ref{surf}. The five lines drawn in each plot
  correspond to different amounts of gas distributed between the
  planets. With increasingly rapid rates of migration, the amounts of
  gas are $1\,\rm M_J$, $3\,\rm M_J$, $5\,\rm M_J$, $10\,\rm M_J$ and
  $20\,\rm M_J$. The upper plot is for the outer planet (moving
  outwards from $10\,\rm AU$) and lower plot for the inner planet
  (moving inwards from $5\,\rm AU$).}  \protect
\label{bet}
\end{figure}

Figure~\ref{bet} shows the inward and outward migration of the two
planets in this case. We also show the effect of varying the amount of
disc mass between the two planets. It is evident that the more mass
there is between the two planets, the faster they migrate, even though
the viscous evolution timescale $\tau_\nu \sim R^2/\nu \propto
R^{2-\beta}$ is a fixed function of radius, independent of surface
density.  The dependence comes about because a higher mass disc exerts
stronger tidal torques on the planets.  In strict Type~II migration,
the migration rate is independent of the disc mass. However, Type~II
conditions do not hold here because there is no disc interior to the
inner planet and the planet mass is comparable to the disc mass.  For
higher mass initial discs, there is more angular momentum in the
system which aids the outer planet's migration to larger distances.

In practice the inner planet would eventually fall into the star. At
an age of $5\times 10^6\,\rm yr$, a $1\,\rm M_{\odot}$ protostar has a
radius of $1.5\,\rm R_\odot$ (Tout, Livio \& Bonnell 1999). Using
their model (Tout, private communication) and equation~(6) of Rasio et
al. (1996) we find that tides in the star capture the planet when it
is about $2.7\,{\rm R_{\odot}}=0.0125\,{\rm AU}$ from the star.  At
this radius, the tides cause the planet to fall into the star faster
than the disc pushes it in. This radius is in line with our choice of
inner disc boundary at 0.01 AU. Once the inner planet has moved in
this far it has surrendered all of its angular momentum and so is no
longer driving any decretion or outward migration of the outer planet.

\subsection{The effect of an inner disc}

\begin{figure}
\epsfxsize=8.4cm 
\epsfbox{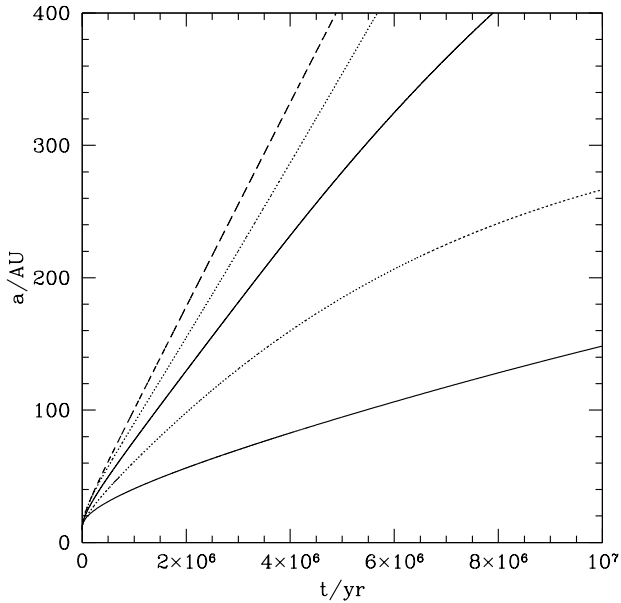}
\caption[] {
  Migration of the outer planet as a function of time.  In each case
  the inner planet starts at $5\,\rm AU$ and the outer planet at
  $10\,\rm AU$.  The fiducial configuration (upper solid line)
  consists of an inner $5 \,\rm M_J$ planet, an outer $1 \,\rm M_J$
  planet and a $\beta =3/2$ gaseous disc consisting of $5\,\rm M_J$
  distributed between the planets. The fiducial configuration is the
  same as the $5 \,\rm M_J$ disc case plotted in Fig~\ref{bet}.  The
  plots described below refer to configurations that consist of a
  single variant on the fiducial case.  The lower solid line has the
  viscosity law with $\beta = 1$.  The dashed line is for a
  configuration with gas interior to the inner planet (see Section
  3.2).  The effect is to speed the outward migration of the outer
  planet and to slow the inward migration of the inner planet.  The
  dotted lines show the effect of changing the mass of the inner
  planet. When the inner planet has a mass of $10 \,\rm M_J$ (upper
  dotted line) the outer planet is forced outwards more rapidly and
  the inner planet moves inwards more slowly. Conversely, when the
  inner planet has a mass of $1 \,\rm M_J$ (lower dotted line), the
  outer planet is forced outwards more slowly and the inner planet
  moves inwards more rapidly.}
\label{ins}
\end{figure}

We now consider the effect of disc material located at radii $R<a_1$,
interior to the inner planet, so that $\Sigma_{0 \rm in }$ is nonzero
in equation~(\ref{densdistr}).  In Figure~\ref{ins} we compare models
having $5\,\rm M_J$ between the planets with and without disc matter
located inside the inner planet.  The migration of planets with no
inner disc is reproduced from Figure~3. For comparison we show the
case (dashed line) where the initial surface density between the
planets is the same but the surface density profile is then continued
inwards to the inner radius with $\Sigma_{0 \rm in }=\Sigma_{0 \rm bet
}$.  For this case with mass interior to the inner planet, the inner
disc mass is $15.9\,\rm M_J$.

We see that the mass interior to the planets causes the outer planet
to migrate outwards faster. The inner planet also migrates inwards
more slowly. This is a consequence of the inner disc providing a
positive torque on the inner planet and an additional source of
angular momentum.

\subsection{Effect of the mass of the inner planet}

In Figure~\ref{ins} we also show the effect of changing the mass of
the inner planet. The two dotted lines show the difference in
migration for the inner planet mass of $1 \,\rm M_J$ versus $10 \,\rm
M_J$. We see that with a more massive inner planet, the outward
migration of the outer planet proceeds on a shorter timescale.  As we
increase the mass of the inner planet, it migrates more slowly towards
the star.  In strict Type~II migration, the migration rate is
independent of the planet mass. However, Type~II conditions do not
hold here because there is no disc interior to the inner planet and
the planet mass is comparable to the disc mass.  The migration rate of
a single planet with similar disc conditions decreases with planet
mass but is accurately described by 1D models for circular orbit
planets with $Re =10^5$ (D'Angelo, Lubow, \& Bate 2006).  The result
is that the gas between the planets moves in more slowly and hence
that the outer planet migrates outwards more rapidly, closer to the
case described in Section~\ref{sec:addec}.

\subsection{Effect of $\beta$}

In Figure~\ref{ins} we also show the effect of varying $\beta$, the
exponent of $R$ in the assumed viscosity law while maintaining a
constant viscosity value at radius $1\,\rm AU$ in
equation~(\ref{viscosity}).  We note that for $\beta = 1$ the
Reynold's number $Re \propto R^{-1/2}$ falls off more slowly with
radius, thus increasing the range of validity of the Type II migration
torque formulae. For $\beta = 1$, the viscosity is smaller
at large radii throughout the fiducial disc in Figure~\ref{ins} and so
the overall viscous timescale $\propto R^2/\nu$ increases.  As a
consequence we see that migration to large radii occurs on a much
longer timescale with smaller $\beta$.

\subsection{Effect of an outer disc}

\begin{figure} 
\epsfxsize=8.4cm
\epsfbox{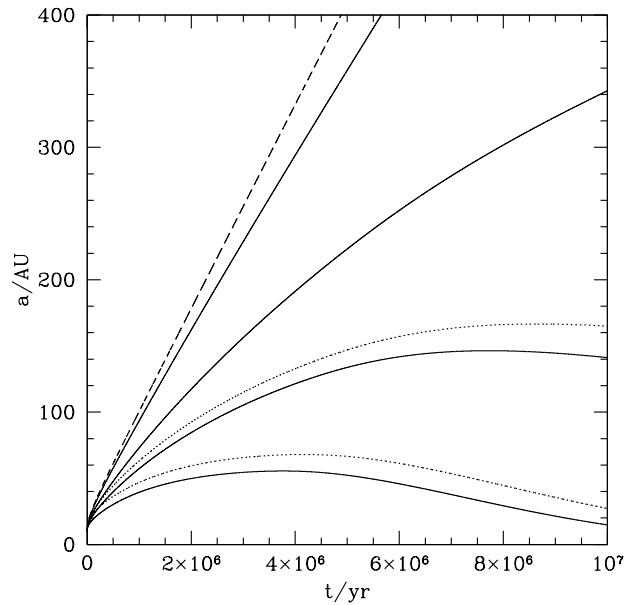}
\caption[] {Migration of the outer planet as a function of time for
  different mass outer discs. Each case consists of a $5\rm \, M_J$
  inner planet that starts at $5\,\rm AU$, a $1\,\rm M_J$ outer planet
  that starts at $10\,\rm AU$, a gaseous disc $5\,\rm M_J$ that is
  distributed between the planets and an inner gaseous disc of
  $15.9\,\rm M_J$.  The amount of gas in the outer disc, exterior is
  the outer planet, is varied in the plots. The gaseous discs have
  $\beta=3/2$. The solid lines plot cases when the outer disc is
  initially truncated at $20\,\rm AU$.  The uppermost solid line
  corresponds to the outer disc containing only $0.1\,\rm M_J$ and is
  almost indistinguishable from the case of there being no mass in the
  outer disc, shown in the dashed line in Figure~\ref{ins} and here.
  As the mass in the outer disc increases with masses of $0.1\,\rm
  M_J$, $0.5\,\rm M_J$, $1\,\rm M_J$ and $2\,\rm M_J$, the inner
  planet migrates inwards more quickly and outward migration of the
  outer planet is increasingly slowed and even reversed.  The dotted
  lines plot cases with $1$ and $2\,\rm M_J$ in the exterior disc but
  with the matter in the outer disc now distributed out to a much
  larger truncation radius of $R_{\rm t}=100\,\rm AU$.  This
  demonstrates that the migration behaviour depends predominantly on
  the total mass in the outer disc, rather than on its distribution
  for a $\beta=3/2$ disc.  }
\label{out}
\end{figure}

Finally in this section, we consider the more realistic case of the
surface density distribution extending both interior to and exterior
to the pair of planets, i.e., $\Sigma_{0 \rm in}$ and $\Sigma_{0 \rm
  out}$ are both nonzero in equation~(\ref{densdistr}).  The disc is
initially truncated at $R_{\rm t}=20\,\rm AU$ unless otherwise
stated. We consider several values for the disc mass exterior to the
outer planet while $5\,\rm M_J$ of disc mass resides between the two
planets and $15.9\,\rm M_J$ interior to the inner planet. The mass
distribution follows equation~(\ref{densdistr}) and the outer disc
mass is varied by changing parameter $\Sigma_{0 \rm out }$.  In
Figure~\ref{out} we plot the migration of the planets for various
values of the outer disc mass.

As we would expect with more mass in the outer disc, the outward
migration of the outer planet is slower and sometimes reverses.  With
$0.1\,\rm M_J$ in the outer disc the migration is not very different
from the case of no outer disc. With $2\,\rm M_J$ in the outer disc,
the planet migration is initially outwards.  However, the torque from
the outer disc is strong enough to reverse the migration back towards
the star. Once the inner planet has fallen into the star, the outer
planet continues to migrate inwards and the disc behaves as a simple
accretion disc.

In each of these cases, there is a jump in surface density (from
$\Sigma_{0 \rm ins }$ to $\Sigma_{0 \rm out }$) across the radius of
the outer planet. If there were no jump then for $R_{\rm t}=20 \,\rm
AU$ and $M_{\rm bet}=5\,\rm M_J$, the outer disc mass would be
$16\,\rm M_J$.  This mass is greater than the outer disc mass for the
plotted models.  For this outer disc mass it is clear that the outer
planet would not migrate out very far before it gets pushed back in by
the large torque exerted by the outer disc.  We also investigated the
case of a smaller disc that had no surface density jump across the
outer planet, This configuration has a $1\,\rm M_J$ outer disc with
initial truncation radius $R_{\rm t}=11.12\,\rm AU$.  The evolution of
this model is not very different from our model truncated at
$R=20\,\rm AU$ with the same mass outer disc.  We conclude that the
outer disc mass has more influence on outer planet migration than the
initial disc truncation radius.  To demonstrate this further, we also
show in Figure~\ref{out} the effect of increasing the truncation
radius of the disc while keeping the outer disc mass fixed.

The torque from the outer disc acting on the planet is determined by
the surface density close to the planet. Suppose we fix the initial
mass of the outer disc but vary its distribution. Because the outer
disc is a decretion disc, within one of its own viscous times it
relaxes to have $\Sigma \propto R^{-(\beta + 1/2)}$. If $\beta > 1/2$,
then the mass is concentrated at the inner edge and so for a given
mass, the surface density close to the planet is roughly the same.

The time the outer disc takes to relax depends on the viscous
timescale of the initial mass distribution. The longest viscous
timescale at that time depends on $R_{\rm t}$ (for $\beta < 2$). What
really matters is the viscous timescale at the radius where most of
the mass is initially. For $\beta > 1$ most of the mass is initially
at the inner edge. So for $\beta > 1$ the initial transient timescale
does not depend on $R_{\rm t}$. Conversely, for $\beta < 1$, the mass
is predominantly initially at $R_{\rm t}$ and so the relevant
timescale does depend on $R_{\rm t}$.

Consider a decretion disc with the $V_{\rm R}=0$ inner boundary
condition.  It can be shown that the torque is given by
Equation~\ref{torque} evaluated at $R=R_{\rm in}$ which follows from
the azimuthal force equation.  The initial surface density for steady
accretion with $\beta=1$ corresponds to a disc mass
\begin{equation} 
M_{\rm disc} = 2 \pi \Sigma_0 R_{\rm in}(R_{\rm t} - R_{\rm in}). 
\end{equation} 
At early times the density is the same as the initial one so that for
fixed $R_{\rm in}$, $G \propto M_{\rm disc}/(R_{\rm t}-R_{\rm in})$ at
early times. At late times $\Sigma$ becomes independent of $R_{\rm t}$
as the disc forgets its initial conditions and approaches the
self-similar solution. The torque becomes independent of $R_{\rm t}$
but remains linear in $M_{\rm disc}$. Hence we see that the migration
of the outer planet depends on the amount of mass outside it and not
the distribution.

\begin{figure} 
\epsfxsize=8.4cm
\epsfbox{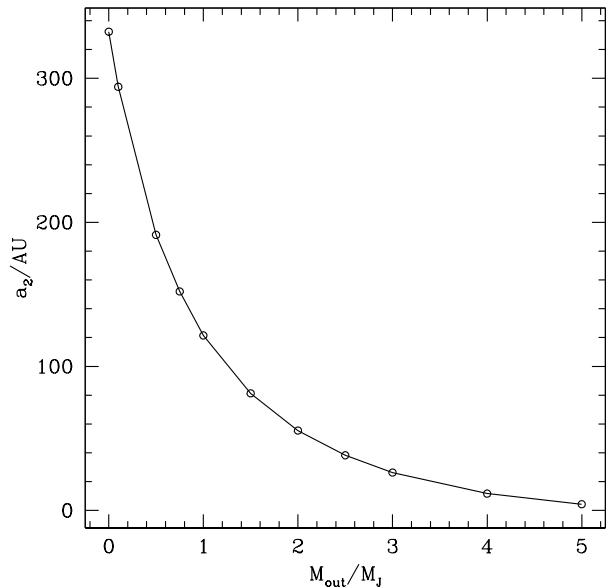}
\caption[]{ The position of the outer planet at a time $t=4\,\rm Myr$
  is plotted against the amount of mass in the exterior disc for the
  cases (solid lines) shown in Figure~\ref{out}. It is evident that in
  order for the outer planet to be able to be driven to large radii,
  the amount of mass in the exterior disc must be less than, or
  comparable to, the mass of the outer planet ($1\,\rm M_J$ in this
  case).}
\label{massout}
\end{figure}

We plot in Figure~\ref{massout} the orbital radius of the outer
planet at a time $t=4\,\rm Myr$ against the exterior disc, $M_{\rm
  out}$. As we remarked before, the lower the mass in the outer disc,
the further the outward migration of the outer planet. For the
particular disc model we discuss here with $\beta=3/2$, in order for
the outer planet to migrate to large radii (say $> 100\,\rm AU$), the
amount of mass remaining in the outer disc after the formation of the
outer planet needs to be less than, or of order, the mass of the
planet.

\subsection{Effect of Planetary Accretion}

Because the gap around the planet is in general not completely
cleared, it is possible for accretion onto the planet to occur. This
effect is not always included in numerical computations of disc
torques. This is partly because high numerical resolution is required
in order to make sure that the details of the flow of material onto
the planet is properly converged, and partly because of uncertainties
about the reaction of the planet to accretion in terms of dissipation
and radiation of accretion energy. Thus in torque computations the
planet is often represented either as a softened potential or as a
sink particle of fixed size.

There are however a number of estimates of accretion rates onto
planets in such discs. These have been considered by (Veras \&
Armitage 2004), who give a fit to the accretion rates derived from
numerical simulations in the form
\begin{equation}
\dot M_{\rm p}=1.668f\dot M_{\rm th}\left(\frac{M_p}{M_J}\right)\exp \left(-\frac{M_{\rm p}}{1.5M_{\rm J}}\right)+0.04,
\end{equation}
where $f$ is a constant parameter which we vary with $0\le f\le 1$ and
$\dot M_{\rm th}=3\pi \nu \Sigma$ is the accretion rate through the
disc further out from the planet.  We consider the effect of applying
this formula to the outer planet. We neglect accretion on to the inner
planet because the inner planet is significantly more massive.

\begin{figure} 
\epsfxsize=8.4cm
\epsfbox{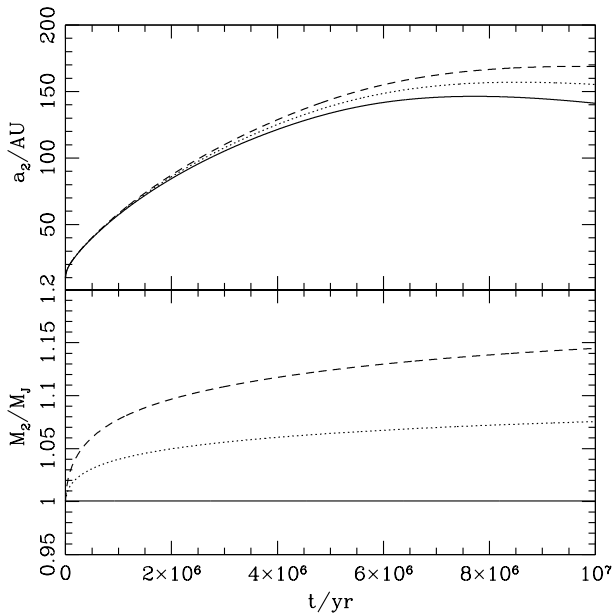}
\caption[] {Migration (upper plot) and mass (lower plot)
  of the outer planet as a function of time for different accretion
  rates on to the outer planet. Each case consists of a $5\rm \, M_J$
  inner planet that starts at $5\,\rm AU$, a $1\,\rm M_J$ outer planet
  that starts at $10\,\rm AU$, a gaseous disc $5\,\rm M_J$ that is
  distributed between the planets and an inner gaseous disc of
  $15.9\,\rm M_J$.  The amount of gas in the outer disc, exterior is
  the outer planet is initially $1\,\rm M_J$ but decreases as mass is
  accreted on to the outer planet. The gaseous discs have $\beta=3/2$.
  The outer disc is initially truncated at $20\,\rm AU$.  The solid
  lines have no accretion on to the planet. The dotted lines have
  $f=0.5$ and the dashed lines have $f=1$.}
\label{planetaccretion}
\end{figure}

We remove the required amount of mass from the first zones with
non-zero mass outside the planet's gap and accrete the mass and
angular momentum of this material on to the planet.  We consider the
case of a $5\rm \, M_J$ inner planet that starts at $5\,\rm AU$, a
$1\,\rm M_J$ outer planet that starts at $10\,\rm AU$, a gaseous disc
$5\,\rm M_J$ that is distributed between the planets and an inner
gaseous disc of $15.9\,\rm M_J$.  The amount of gas in the outer disc,
exterior is the outer planet is $1\,\rm M_J$.  We run models with
$f=0$,~$0.5$ and~$1$. We see in figure~(\ref{planetaccretion}) that
the accretion has little effect on the planetary migration, in fact,
the planets migrate further out with this accretion. There are two
effects here. First, the outer planet extracts angular momentum from
the outer disc and so moves out more quickly. Second, the mass of the
outer disc is steadily reduced. We conclude that if the Veras \&
Armitage (2004) formula is correct, then the effect accretion is small
and all our previous migration results would not be significantly
affected had we included accretion on to the outer planet.

\section{Effect of photoevaporation }
\label{sec:photo}

We have seen in the previous section that outward migration to large
radii is possible only if the outer disc, at radii $R>a_2$, is for
some reason depleted of gas. One means of doing this which we
investigate here, is by photoevaporation.

The surface layers of a disc can be heated by the central star to
sufficiently high temperature that the gas can escape the gravity of
the star. Shu, Johnstone \& Hollenbach (1993) suggested that
protoplanetary discs are dispersed because of heating by ultraviolet
radiation from the central star. The ionizing flux creates a
photoionized disc atmosphere where the gas can become hot enough to
be gravitationally unbound and escape from the system. The thermal
energy of the gas is greater than the gravitational binding energy
beyond the gravitational radius
\begin{equation}
R_{\rm G}=\frac{GM_\star}{c_s^2} \approx 10 \left(\frac{M_\star}{M_\odot} \right){\rm AU},
\end{equation}
where $c_s$ is the speed of sound in the disc's HII atmosphere at $10^4\,$K.

\subsection{Photoevaporation only in $R\ge R_{\rm G}$}

\begin{figure} 
\epsfxsize=8.4cm 
\epsfbox{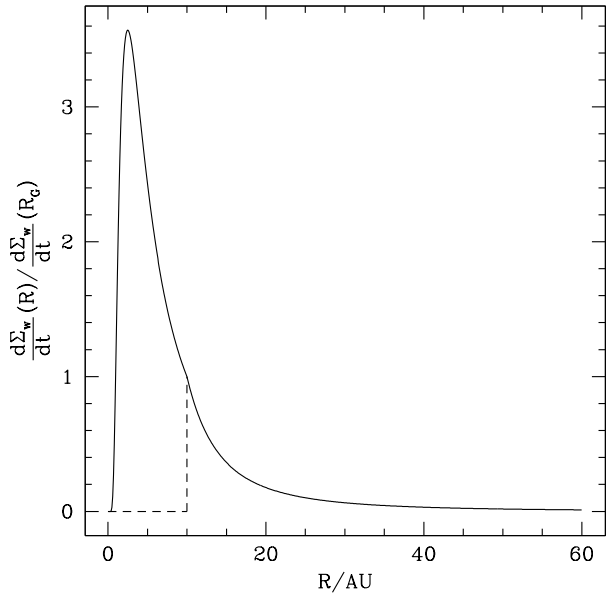} 
\caption[] {The rate of change of surface density scaled to the rate
  at $R_{\rm G}$ is plotted against radius. The dotted line
  corresponds to photoevaporation only in the range in $R\ge R_{\rm
    G}$, as given by equation~(\ref{Sigmaw}). The solid line takes
  account of possible outflow from within that radius and is given by
  equation~(\ref{Sigmaw1}). }
\label{photoevap}
\end{figure}

Hollenbach et al. (1994) used detailed models of the density at the
base of the photoevaporating atmosphere and found a wind mass flux as
a function of radius given by
\begin{equation}
\label{Sigmaw}
\dot \Sigma_{\rm w} = \begin{cases}0 & \text{if $R \le R_{\rm G}$}, \cr
\dot \Sigma_0 \left(\displaystyle\frac{R}{R_{\rm G}}\right)^{-5/2} &  \text{if $R \ge R_{\rm G}$} ,\end{cases}
\end{equation}
 where
\begin{equation}
\dot \Sigma_0 =1.16\times 10^{-11}\left(\frac{\phi}{10^{41}\rm s^{-1}}\right)^{1/2}
\left(\frac{R_{\rm G}}{\rm AU} \right)^{-3/2}\rm M_\odot\, AU^{-2}\, yr^{-1}
\end{equation}
and $\phi$ is the rate of ionizing photons coming from the star. This
rate of change of surface density is plotted against the distance from
the star in Figure~\ref{photoevap} as the dashed line.  Alexander,
Clarke \& Pringle (2005) analysed emission measures and found rates of
ionizing photons from the chromospheres of five classical TTs in the
range $10^{41}-10^{44}\,\rm photon \, s^{-1}$.  We calculate the total
wind mass loss rate from the disc to be
\begin{align} \notag
\dot M_{\rm w} &= \int_0^\infty2\pi R \dot \Sigma_{\rm w}\, dR = 4\pi \dot \Sigma_0 R_{\rm G}^2 \\ 
& = 1.45\times10^{-10}\left( \frac{\phi}{\rm 10^{41} s^{-1}}\right)^{1/2}\left(\frac{R_{\rm G}}{\rm AU}\right)^{1/2}\,\rm M_\odot\,yr^{-1} .
\end{align}

\begin{figure} 
\epsfxsize=8.4cm
\epsfbox{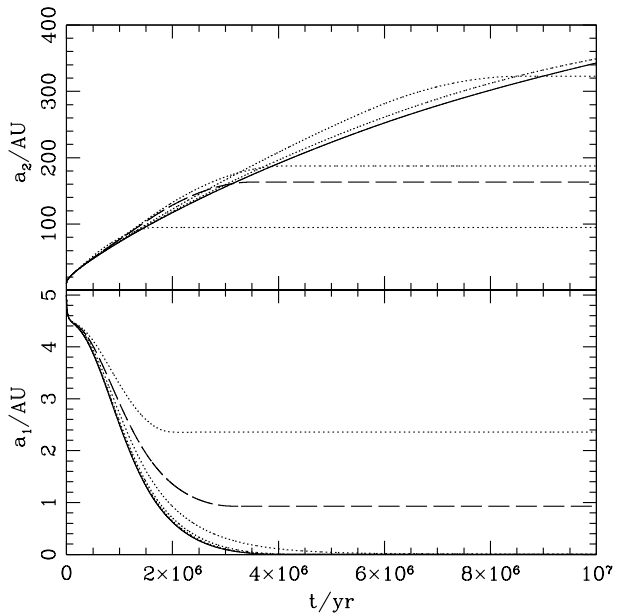}
\caption[] {Migration of the planets shown as a function of
  time accounting for the effects of photoevaporation.  The upper plot
  is for the outer planet and the lower plot for the inner planet. The
  $5\,\rm M_J$ planet starts at $5\,\rm AU$ and the $1\,\rm M_J$
  planet at 10 AU.  There is initially a gas disc of mass of $5\,\rm
  M_J$ between the planets, an inner disc of mass $15.9\,\rm M_J$ and
  an outer disc of $0.5\,\rm M_J$ which is truncated at $20\,\rm AU$.
  The dotted lines show the effects of photoevaporation according to
  equation~(\ref{Sigmaw}). The solid lines are identical to the
  $0.5\,\rm M_J$ exterior disc case shown in Figure~\ref{out} without
  photoevaporation, $\phi=0$.  The line corresponding to the lowest
  photon rate of $\phi = 10^{40}\,\rm s^{-1}$ lies above solid the
  line for $\phi = 0$. This is because photoionization in this case is
  able to remove some of the mass in the outer disc and so enables
  faster outward migration.  The line corresponding to $\phi =
  10^{41}\,\rm s^{-1}$ lies above the $\phi = 0$ line for some time
  initially for the same reason but eventually photoevaporation
  removes enough disc material from between the two planets to cut
  communication and so to halt further outward migration. The same
  applies to the lines corresponding to $\phi= 10^{42} \rm \,s^{-1}$
  and $10^{43}\rm \,s^{-1}$ with the halt in migration occurring
  sooner for the larger photoevaporation rate.  The dashed lines use
  photoevaporation model described in equation~(\ref{Sigmaw1}) with
  $\phi=10^{42} \rm \,s^{-1}$.  }
\label{photo0p5}
\end{figure}

We now consider the evolution of an initial setup as described in
Section~3.5 with a gas disc of mass of $5\,\rm M_J$ between the
planets, an inner disc of mass $15.9\,\rm M_J$ and an outer disc of
$0.5\,\rm M_J$ but now allow for disc depletion according to the
formula given in equation~(\ref{Sigmaw}) for various values of the
photon flux $\phi$.  The results are shown in Figure~\ref{photo0p5}.
The migration initially proceeds on a faster timescale than with no
mass loss by photoevaporation. This is because initially most of the
mass lost arises from the outer disc. This decrease in mass means
the negative torque on the planets is smaller and so they migrate
outwards faster.  However, once the outer planet has moved
significantly outwards, the main effect of photoevaporation is to
remove the gas from between the planets. Once sufficient gas has been
removed, all contact is lost between the planets and the inner planet
is no longer able to give up angular momentum to the outer one and
migration stops.

\begin{figure} 
\epsfxsize=8.4cm
\epsfbox{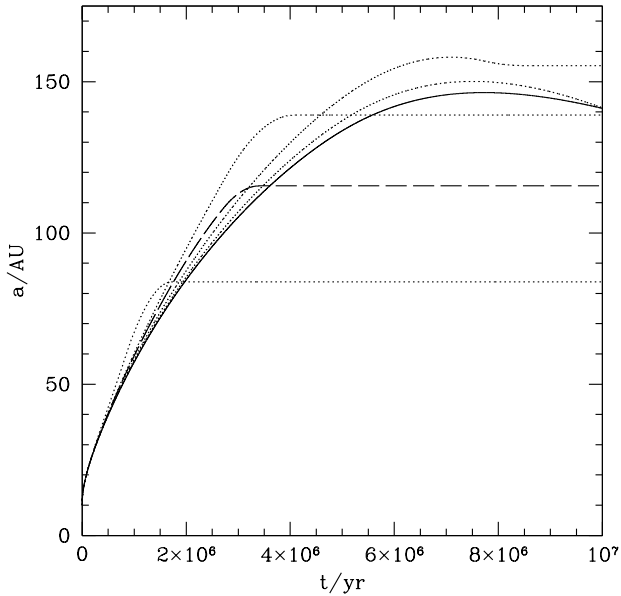} 
\caption[] { This plot describes the same configuration as
  Figure~\ref{photo0p5} except that here there is a higher initial
  outer disc mass $1.0\,\rm M_J$. We plot the migration of the outer
  planet.  As before, the solid line corresponds to the case $\phi=0$
  shown in Figure~\ref{out}.  The dotted line corresponding to $\phi =
  10^{40} \rm \,s^{-1}$ lies at all times just above the $\phi = 0$
  line for the same reason as discussed in Figure~\ref{photo0p5}. The
  other dotted lines correspond, in order of decreasing final radius
  in this diagram, to $\phi = 10^{41}, 10^{42}, 10^{43}\rm \,s^{-1}$.
  By comparison with Figure~\ref{photo0p5}, we see that for a given
  photon flux, or photoevaporation rate, a larger initial mass in the
  exterior disc leads typically to a larger final radius for the
  outwardly migrating planet.  The dashed lines have the
  photoevaporation model described in equation~(\ref{Sigmaw1}) with
  $\phi=10^{42} \rm \,s^{-1}$.}
\label{photo1}
\end{figure}

In Figure~\ref{photo1} we show the results of an identical set of
computations, except that initially the amount of gas in the outer
disc is increased to $1\,\rm M_J$.  With stronger photoevaporation,
the distance the outer planet moves decreases and its migration
ceases due to disc dispersal.

\subsection{Varying the radial dependence of the evaporation rate.}

There have been more recent suggestions that photoevaporation happens
as far in as $0.1-0.2\,R_{\rm G}$ (Liffman 2003, Adams et al. 2004,
Font et al. 2004).  To simulate the effects of this form of
photoevaporation, we have taken the model from the previous section
and changed the gravitation radius to $R_{\rm G}=1\,\rm AU$ with the
same range of $\phi$. In this case, we find the photoevaporation has
little effect on the disc.  This is because now photoevaporation
mostly removes mass from radii well within the planetary orbits.  We
have already seen in Section~3.3 that the presence or absence of an
inner disc has only a limited effect on the migration.

Alternatively, in line with the ideas of Dullemond et al. (2006) who
model the photoevaporation as a Bernoulli flow, we take the radial
dependence of the rate of change of surface density due to
photoevaporation as
\begin{equation} 
\label{Sigmaw1}
\dot \Sigma_{\rm w}(R)=\begin{cases}\dot \Sigma_0\displaystyle \exp \left({\frac{1}{2}\left(1-\frac{R_{\rm G}}{R}\right)}\right)
  \left(\displaystyle\frac{R}{R_{\rm G}}\right)^{-2} &\text{if $R \le R_{\rm G}$}, \cr
\dot \Sigma_0 \left(\displaystyle\frac{R}{R_{\rm G}}\right)^{-5/2} &  \text{if $R \ge R_{\rm G}$} .\end{cases}
\end{equation}
The spatial form is the same as equation~(\ref{Sigmaw}) for $R\ge
R_{\rm G}$ but extends further inwards to smaller radii.  We show
this graphically as the solid line in Figure~\ref{photoevap}. We see
that in this case, the peak surface density loss occurs at $0.25
R_{\rm G}$ whereas for equation~ (\ref{Sigmaw}) the peak loss is
occurs at $ R_{\rm G}$.

The dotted and dashed lines in Figure~\ref{photo0p5} show the
migration of planets with discs undergoing photoevaporation described
by equation~(\ref{Sigmaw}) and equation~(\ref{Sigmaw1}), respectively.
We see that with the inner photoevaporation described by
equation~(\ref{Sigmaw1}), the outer planet does not migrate out as
far. This is because the total rate of loss of mass is increased and
so the disc is removed on a shorter timescale.  The peak in $\dot
\Sigma _{\rm w}$ is now inside of the inner planet.  Mass removed from
the inner disc has little effect on the migration of the planets (see
Figure~\ref{ins}).

\section{Observed Systems}

We now consider whether the mechanism discussed here for moving planets out
to large distances from the star can explain the structure of systems
inferred to have planets at $R>10\,\rm AU$.  Such systems have been
discovered through direct imaging (Chauvin et al.  2004) and interpreted in
terms of planet perturbations on dust discs (Wyatt et al. 1999).

Planets discovered through direct imaging have a relatively high mass ratio.
For example, the system 2MASS 1207334-393254 (2M1207) has a $5\,\rm M_J$
planet at $55\,\rm AU$ from a $25\,\rm M_J$ brown dwarf (Chauvin et al.
2004). The mass ratio is closer to those in binary stars than for known star
planet systems (Lodato et al. 2005).  Lodato et al. (2005) argue that the
planet could have formed {\it in situ} by gravitational instability but not
through core accretion owing to the prohibitively long planet formation
timescales at this distance.  The planet could however have formed in a
reasonable time at $0.6\,\rm AU$ (Lodato et al.  2005).  They suggest that
this is too near to the star for outward migration to provide the current
separation.  Using our simple model and equation~(\ref{simple}) we find that
to drive out the planet to 55 AU the inner planet would have to be comparable
in mass to the star 2M1207. In other words, for this scenario to work we
would require 2M1207 to have, or to have had, a close binary companion, and
the $5\,\rm M_J$ planet must have formed in a circumbinary disc. Assuming the
companion is of similar mass to 2M1207, equation~(\ref{simple}) implies that
the current planet parameters are compatible with a stellar companion
originally at 1.3 AU, and a formation location for the planet of 2.9AU.

The inferred debris-disc planets are found in two different types of systems,
old systems and young systems.  Most are found in relatively old systems
(much older than $10\,\rm Myr$) and inferred to lie close to the inner edge
of a planetesimal belt at greater than $30\,\rm AU$. No gas is seen in these
systems (Dent et al. 2005) and the inner regions are devoid of dust and
planetesimals (Wyatt 2005b).  The existence of planets are inferred from
asymmetries in the structure of the dust discs which have been imaged for the
closest systems.  This type of system is typified by the debris disc around
$350\,\rm Myr$ old Vega. The disc's clumpy structure (Holland et al. 1998,
Wilner et al.  2002) has been used to infer the presence of a Neptune-mass
planet currently located at $65\,\rm AU$ from the star, since that structure
can be explained by the planet having migrated outward from $40\,\rm AU$ over
$56\,\rm Myr$ while trapping planetesimals at its resonances (Wyatt 2003,
Wyatt 2006). The modelling also permitted different planet masses to have
caused the same clumpy resonant structure as that observed, as long as the
migration rate is changed accordingly.

It seems reasonable to assume that the migration scenario proposed here could
explain distant planets in systems like Vega, since, if their existence is
confirmed, the outward migration of such planets would naturally explain both
their large orbital radii and the clumpy structure of their debris discs.
For this scenario to work, there are certain requirements on the outer planet:
\begin{list}{}{}
\item (i) the migration must occur before the gas disc dissipates for
  which we require a migration timescale of less than $6\,\rm Myr$
  (Haisch, Lada \& Lada 2001), which means the planet mass must be greater
  than $0.25\,\rm M_J$ to cause the observed clumpy dust structure (Wyatt
  2003);
\item (ii) to migrate by Type II migration at $40\,\rm AU$ the planet
  must be greater than roughly $1\,\rm M_J$ in order to open a gap in
  the disc (Bryden et al 1999);
\item (iii) to have avoided detection in direct imaging surveys the
  planet must have a mass less than $7\,\rm M_J$ (Macintosh et al.
  2003, Hinz et al. 2006).
\end{list}
Thus we consider that a $2\,\rm M_J$ planet that migrated from $40$ to
$65\,\rm AU$ over $0.3\,\rm Myr$ could explain the observed disc structure,
parameters which can be reproduced with the migration mechanism proposed
here. The planet could also have started closer in if there is some mechanism
to remove material from the resonances at high eccentricity.

This interpretation would make the prediction that there was at one time (and
possibly still is if it has not already been accreted) a more massive planet
close to Vega. The fact that this system is close to pole-on (Aufdenberg et
al. 2006) indicates that such a planet would be hard to detect by radial
velocity measurements.

However, we cannot escape from the requirement of the proposed model that the
outer disc be deficient in mass, in order to provide outward migration. If
the deficiency were due to truncation, then the truncation would have to
apply to the gas disc and not the solids. The reason is that clumpy structure
in the debris disc is modelled as planetesimals that were present during the
migration process, and which extended beyond the 2:1 resonance of the
outer planet at $1.59 a_2$. This makes problematic the possibility that
the deficiency might have been caused by physical truncation due to
stellar encounters in a high density star formation environment. However,
photoevaporation of gas remains a possible mechanism.

Planets have also been inferred from structures in young $5-20\,\rm
Myr$ systems. Such systems have both dust and gas at a wide range of
distances.  These are typified by $5\,\rm Myr$ old HD141569, for which
tightly wound spiral structure at $325\,\rm AU$ (Clampin et al.  2003)
has been used to infer the presence of a planet of $0.2-2\,\rm M_J$ at
$235-250\,\rm AU$ (Wyatt 2005b).  Spiral structure in a ring at
$185\,\rm AU$ (Clampin et al.  2003) may also be indicative of a
Saturn mass planet at $150\,\rm AU$ (Wyatt 2005b). Like the planet in
2M1207, the planets in this system are unlikely to have formed {\it in
  situ} unless by gravitational instability. However, the migration
scenario proposed here provides a viable mechanism through which these
planets could migrate out to their current locations following
formation closer to the star, since an inner planet in this scenario
could equally drive out more than one planet. If this interpretation
is correct, then we would make the prediction that a more massive
planet exists, or existed, at a distance of less than a few AU from
HD141569. The migration must also have occurred on timescales much
shorter than 5 Myr, since some time is required after the migration
for the spiral structure to be imprinted on the disc (Wyatt 2005b).

The current distributions of gas and dust in this system provide some clues
as to whether this scenario is feasible. There certainly appears to be a
radially extended gas distribution which might in the past have contained
enough mass interior to the planet at 150 AU to cause rapid outward migration
(Jonkheid et al. 2006), and there is no evidence that the mass exterior to
the planets would have been sufficient to prevent that migration. However,
there are a number of density variations in the radial distributions of gas
and dust which would be hard to explain within the context of this model
(Marsh et al. 2002; Merin et al. 2004; Goto et al. 2006; Jonkheid et al.
2006). Such variations are likely caused by processes not included in the
current scenario, such as the dynamical interaction between gas and dust
(e.g., Takeuchi \& Artymowicz 2001, Krauss \& Wurm 2005), photoevaporation of
gas near the gravitational radius (Clarke, Gendrin \& Sotomayer 2001), and
grain growth which may have occurred after the planetary migration.

\section{Conclusions}

We have shown that it is possible {\it in principle} for planets to
form at small radii and then to migrate out to large radii from their
star by means of planet-disc interactions.  The possible mechanism we
discuss here involves the formation of a massive inner planet and of a
less massive outer planet suitably spaced in radius that there is
enough gas between them to effect angular momentum transfer and
suitably spaced in time that the first to form has not migrated too
far before the formation of the second.  In view of the uncertainties
surrounding the planet formation process, we have not here addressed
the plausibility of setting up such a configuration. 

We also require that tidal torques act efficiently on the planet, even
at large radii. This leads to a conflict of requirements on the size
of the disc viscosity in that a large outward torque requires the
opening of a clean gap and so a small viscosity (e.g. Crida \&
Morbidelli 2007), whereas too small a viscosity implies that the
migration takes too long and cannot take place before the disc has
been dispersed. Whether there is a finite range of viscosity between
these two constraints depends on the detailed properties of such
discs, and in particular on the unknown nature and magnitude of the
disc viscosity (e.g. King, Pringle \& Livio, 2007)

To compute the evolution of disc and planetary orbits we have used a
1D disc approximation, used an idealised model of the disc structure,
and and have used a standard formulation which attempts to estimate
the interactive disc-planet torques, under the assumption that the
planets are massive enough to open a clean gap. In reality it would be
better, but much more computer-intensive, to undertake the disc/planet
evolution while the torques are being computed using 2D, or better 3D,
hydrodynamics simulations for the gas in the neighbourhood of the
planet (c.f. Crida et al., 2007), and to solve simultaneously for the
local disc structure (for example assuming thermodynamic equilibrium,
e.g. Bell et al., 1997) and in addition taking account of heating of
the disc by the central star (e.g. Garaud \& Lin, 2007). But even so,
there still remain sufficient uncertainties about the basic physics
involved that it is not always easy to assess the degree to which such
simulations reflect physical reality. Even within a given set of
simulations, with a fixed set of assumptions, it is necessary to take
some trouble to ensure that the calculations have enough resolution
that the torque estimates have converged (see, for example,
D'Angelo, Bate \& Lubow, 2005). Although we have made an attempt to
estimate its effects, it is clear that the degree, nature and effects
of accretion onto the planet have yet to be fully understood. And in
addition, almost all simulations so far assume that the disc viscosity
is some kind of fixed form of Navier-Stokes viscosity, whereas in
reality the viscosity is most likely due to some form of
magneto-hydrodynamics turbulence (e.g. Nelson \& Papaloizou, 2004),
and may in the outer, cooler parts of the disc be spatially confined
to a small part of the disc (Gammie, 1996).

Once the planet formation is complete, we find that for significant
outward migration to occur, it is necessary for there to be very
little or no disc mass exterior to the outer planet: the lower the
surface density exterior to the outer planet, the faster and further
the migration.  One possible way of achieving this is through
dynamical truncation of the disc, either by a fly-by in a dense
environment or through the presence of a distant binary companion or
through a multi-body interaction during the formation process (Clarke
\& Pringle 1991). We have investigated the possibility that the outer
disc is reduced by photoevaporation. If the source of photoevaporation
is external, due for example to the nearby presence of a hot star,
then the disc can be evaporated from the outside as we require (and as
modelled in effect by Veras \& Armitage, 2004). In this paper we have
investigated the possibility of photoevaporation being caused by the
central star. We find that, since such photoevaporation depletes the
disc predominantly at around $R_G \approx 10\,\rm AU$, the general
effect is in this case to hinder rather than enhance migration to
large radii.

The main conclusion from these calculations is that while outward
migration is indeed a feasible mechanism for the production of gas
giants at large distance form the central star, the parameter space in
which its occurrence is possible may, for the reasons given above, be
somewhat limited.

Outward planet migration is also possible in circumbinary discs. We
consider the case that the binary's angular momentum is at least
comparable to that of the surrounding disc.  Planets which open gaps
undergo outward migration in such discs, such as shown in
Figure~\ref{trace}.  The formation of the planet can occur after the
formation of the binary while the disc is present.  Unlike the case
involving an inner planet discussed above, the disc mass external to
the planet need not be small. The main issue is whether the disc
behaves as a decretion disc.  Two-dimensional simulations indicate
that some material can accrete past the disc inner edge and onto the
binary (Artymowicz \& Lubow 1996; Gunther \& Kley 2002). The radial
distribution of planets around binary stars could be quite different
from the single star case.

We have discussed briefly the possibility of applying these ideas to
observed systems. The major uncertainties in making such applications
are the timescales on which planets form, the radial positions at
which they form and the timescales on which gaseous discs evolve and
are dissipated. In this paper we have used a simple and idealised
description of the viscosity whereas in reality the viscosity and in
consequence the migration timescales are functions of the local disc
properties as they evolve. Nevertheless we note that at least some of
the observed systems might be compatible with the migration scenario
discussed here.

\section{Acknowledgements}

RGM thanks Christopher Tout for useful conversations.  SHL
acknowledges support from NASA Origins of Solar Systems grant
NNG04GG50G.  JEP thanks the Space Telescope Science Institute for
continuing support under its Visitors' Program.

\end{document}